\documentclass[aps,preprintnumbers,groupedaddress,twocolumn,floatfix,10pt]{revtex4-1}            

\pdfoutput=1

\usepackage{amsmath}
\usepackage{amsfonts}
\usepackage{amssymb}
\usepackage{graphicx}

\newcommand{\eq}[1]{\begin{equation}
                     \begin{split} #1 \end{split}
                     \end{equation}}
\newcommand{\ov}{\overline}
\newcommand{\op}{\hspace{1pt}}

\begin{document}


\preprint{MPP-2015-38, LMU-ASC-10/15, DFPD-2015-TH-7}
 
\title{Towards Axionic Starobinsky-like Inflation in String Theory}

\author{Ralph Blumenhagen$^{1}$} 
\author{Anamar\'{\i}a Font$^{1,2}$}
\altaffiliation{On leave from Dept. de F\'{\i}sica, Fac. de Ciencias, UCV, Caracas}
\author{Michael Fuchs$^{1}$} 
\author{Daniela Herschmann$^{1}$} 
\author{Erik Plauschinn$^{3,4}$}

\affiliation{  
\phantom{aa}\\
$^{1}$Max-Planck-Institut f\"ur Physik, F\"ohringer Ring 6, 80805 M\"unchen, Germany\\
\vskip-9pt
$^{2}$Arnold Sommerfeld Center for Theoretical Physics,
LMU, Theresienstr.~37, 80333 M\"unchen, Germany \\
\vskip-9pt
$^{3}$Dipartimento di Fisica e Astronomia ``Galileo Galilei'',
Universit\`a  di Padova, Via Marzolo 8, 35131 Padova, Italy\\
\vskip-9pt
$^{4}$INFN, Sezione di Padova, Via Marzolo 8, 35131 Padova, Italy
}

 
\begin{abstract}
It is shown that Starobinsky-like potentials  can be  realized
in non-geometric flux compactifications of string theory, 
where the inflaton involves an axion  whose shift symmetry can protect
UV-corrections to the scalar potential.
For that purpose  we evaluate the backreacted, uplifted  F-term axion-monodromy
potential, which interpolates between a quadratic and
a Starobinsky-like form. Limitations due to the requirements 
of having a controlled approximation of the UV theory and
of realizing single-field inflation are discussed. 
\end{abstract}



\maketitle



\section{Introduction}
\label{sec:intro}

The recent release of the PLANCK~2015 data provides improved experimental
results and bounds on the $\Lambda$CDM
cosmology \cite{Ade:2015lrj}.
In particular, the \mbox{BICEP2} observation \cite{Ade:2014xna} of a tensor-to-scalar
ratio as big as $r=0.2$ can now be completely explained by a foreground
dust contamination of the signal, and is replaced by the upper bound 
\mbox{$r<0.113$}. Moreover, for  the spectral index PLANCK~2015 reports $n_s=0.9667\pm
0.004$ and for its running $\alpha_{s}=-0.002\pm 0.013$.

As a consequence, large-field inflationary potentials of the type
$V\sim\Theta^p$ are essentially ruled out for $p\ge 2$,
and the currently best class of 
models fitting the data is plateau-like \cite{Martin:2015dha}. 
This class contains the Starobinsky model \cite{Starobinsky:1980te},
as well as more general Starobinsky-like models
\eq{
\label{starolike}
V(\Theta)= M^4_{\rm Pl} \Bigl(A-B\op e^{-\gamma \op \Theta}\Bigr)\,,
}
(see also \cite{Goncharov:1985yu}, 
as well as \cite{Linde:2014hfa} for a historical perspective on the Starobinsky model).
Starobinsky-like models have been constructed in 
string theory in the LARGE volume scenario (LVS), where the role of
the inflaton is played by a canonically normalized K\"ahler modulus 
\cite{Cicoli:2008gp,Burgess:2013sla}.

When working with a model of large-field inflation, Planck suppressed 
higher-order operators need to be controlled, since otherwise they lead to an $\eta$-problem.  
For the LVS, corrections are suppressed by an exponentially large volume,
while in the case of the inflaton being an axion, the shift symmetry of the latter can 
protect the potential against perturbative corrections. 
Various scenarios for axion inflation have been constructed, such as
natural inflation~\cite{Freese:1990rb}, \mbox{N-flation~\cite{Dimopoulos:2005ac}}, 
or aligned inflation~\cite{Kim:2004rp}.

Another promising string-theoretic  approach,
still allowing for some control over the higher-order corrections, is
axion monodromy inflation \cite{Silverstein:2008sg,McAllister:2008hb},
for which a field theory version has been proposed in 
\cite{Kaloper:2008fb}
(for a review see \cite{Baumann:2014nda}).
In \cite{Marchesano:2014mla,Blumenhagen:2014gta,Hebecker:2014eua} this scenario has been 
realized via the F-term 
scalar potential induced by background fluxes, which
has the advantage that supersymmetry is broken
spontaneously by the very same
effect by which usually moduli are stabilized.  Such models 
were studied in \cite{Blumenhagen:2014nba,Hebecker:2014kva} 
for the possibility to provide a quadratic potential for the axion.

In this letter, we analyze a toy model for the flux-in\-duced scalar potential for
a  large excursion of a prospective axion/inflaton.
For concreteness, we consider a simple type IIB superstring flux
compactification,  where a superpotential for all
moduli is generated by turning on NS-NS and R-R three-form fluxes as
well as non-geometric fluxes.
Following \cite{Dong:2010in}, taking the backreaction of the stabilized moduli onto the evolution of
the inflaton into account, we find the expected flattening of the
uplifted potential, which after canonical normalization 
interpolates between a quadratic and a
Starobinsky-like form. 
Here we discuss 
the cosmological consequences of this model, whereas more details on the formal framework
and on the phenomenology 
will be discussed elsewhere
\cite{Blumenhagen:new}.


\section{Large-field  inflation}

Let us recall the expressions of the cosmological parameters for 
the large-field polynomial and Starobinsky-like inflationary models.
For polynomial inflation with $V\sim \Theta^p$, the slow-roll parameters $\epsilon={1\over 2} \bigl({V'\over V}\bigr){}^2$
and $\eta={V''\over V}$ can be written in the following way
\eq{
   \epsilon = {1\over 2}\, {p^2\over \Theta^2}\,,\hspace{40pt}
   \eta= {p\op (p-1)\over \Theta^2}\, ,
}
and the number of e-foldings is expressed as 
\eq{
   N_e&=\int^{\Theta_*}_{\Theta_{\rm end}} {V\over V'}\,
   d\Theta= {1\over p} \int^{\Theta_*}_{\Theta_{\rm end}}
      \Theta\, d\Theta
   \simeq  {\Theta_*^2\over 2\op p} \, .
}
This implies that $N_e\simeq {p\over 4\epsilon}$,
and the spectral index, its running and the tensor-to-scalar ratio
are obtained as
\eq{
\label{polyvalues}
    &n_s=1+2\op\eta-6\op\epsilon \sim 1-{p+2\over 2 N_e}\,,\\
    &\alpha_s=-{(p-1)(9\op p-14)\over 2\op N_e^2}\,,\hspace{40pt}
    r=16\op \epsilon\sim {4p\over N_e}\, .
}
For the Starobinsky-like model \eqref{starolike}, the slow-roll parameters become
\eq{
\label{staroa}
     \epsilon={1\over 2\op\gamma^2}\op \eta^2\,,
     \hspace{30pt}
     \eta=-{1\over N_e} \,,
}
so that
\eq{
\label{starob}
    n_s=1-{2\over N_e}\,,\hspace{20pt}
    \alpha_{s}=-{2\over N_e^2}\,,\hspace{20pt}
    r={8\over (\gamma \op N_e)^2}\,.
}
Independently of the parameters $A$, $B$ and $\gamma$, for $N_e=60$ e-foldings
this gives the experimental value $n_s\sim 0.967$. Note that $n_s$ and $\alpha_s$ in \eqref{starob}
agree with the values for a quadratic potential  in
\eqref{polyvalues}, except that the tensor-to-scalar ratio  comes out
smaller.

The amplitude of the scalar power spectrum takes the experimental value 
${\cal P}=(2.142\pm 0.049)\cdot 10^{-9}$, and can be
expressed as 
\eq{
\label{powerspec}
   {\cal P}\sim {H^2_{\rm inf}\over 8\op \pi^2\op\epsilon\op M^2_{\rm Pl}} \,.
}
From this one can extract the  Hubble constant during inflation, and
consequently
the mass of the inflaton via $M^2_{\Theta}=3\op \eta\op H^2$.
The relation $V_{\rm inf}=3\op M_{\rm Pl}^2\op H^2_{\inf}$ then fixes  
the mass scale of inflation.


\section{Fluxes and moduli stabilization}

We now turn to the framework of type IIB orientifolds on Calabi-Yau (CY)
threefolds, equipped with geometric and non-geometric fluxes.
The NS-NS and R-R fluxes $H_3$ and $F_3$ generate a
potential for the complex-structure and axio-dilaton moduli, where the
latter is written as $S=s+i\op c$ with $s=\exp(-\phi)$ and $c$ denoting the R-R
zero-form.
Non-geometric $Q$-fluxes can generate 
a tree-level potential for the K\"ahler moduli $T_\alpha=\tau_\alpha+i
\rho_\alpha$, where $\tau_\alpha$ denotes a four-cycle volume (in Einstein frame) and 
$\rho_\alpha$ is the R-R four-form reduced on that cycle.
The details for such flux compactifications have been worked
out in
\cite{Shelton:2005cf,Aldazabal:2006up,Grana:2006hr,Micu:2007rd}
(see also \cite{Hassler:2014mla}).
The resulting scalar potential reads
\eq{
   V={M_{\rm Pl}^4\over 4\pi} \, e^K \left( K^{i\ov j} D_i W D_{\ov j} \ov W -3\op
   |W|^2\right) ,
}
which is computed from  the K\"ahler potential 
\eq{
  \label{k_pot}
  \textstyle 
 K=-\log\Bigl( -i\int \Omega\wedge \ov{\Omega}\Bigr)-\log\bigl(S+\ov S\bigr)
     -2\log {\cal V} \,
}
and the flux-induced superpotential
\eq{
\label{thebigW}
  W=&-\bigl( {\mathfrak f}_\lambda  X^\lambda -\tilde {\mathfrak f}^\lambda  F_\lambda  \bigr) 
  +i\op S \big( h_\lambda  X^\lambda - \tilde h^\lambda  F_\lambda \bigr) \\
  &+i \op T_{\alpha}  \bigl( q_\lambda{}^\alpha   X^\lambda - \tilde q^\lambda{}^\alpha  F_\lambda\bigr)\,.
 }
Note that here ${\cal V}$ denotes the volume of the Calabi-Yau manifold (in Einstein frame), and 
that we have
assumed a large-volume and small string-coupling
regime so that higher-order corrections can safely be ignored. 
As usual, $X^\lambda$ and $F_\lambda$ denote the periods of the holomorphic
three-form $\Omega$, and $\{\op\mathfrak f\op,\op \tilde{\mathfrak f}\op\}$, $\{h,\tilde h\}$ and $\{q,\tilde q\}$
denote the flux quanta of $F_3$, $H_3$ and $Q$.

To be more specific, let us consider a  simple case of a CY manifold
with no complex-structure moduli 
and just one K\"ahler modulus. One might think of it
as an isotropic six-torus with fixed complex structure.
In this case the K\"ahler potential is  given by
\eq{
         K=-3 \log ( T+\ov T )-\log (S+\ov S)\,.
}
Next, we turn on fluxes to generate the  superpotential
\eq{
         W=-i\op  \tilde {\mathfrak f}+i\op  h\op S   + i\op q\op T\, ,
}
with $\tilde{\mathfrak f},h,q\in \mathbb Z$. 
The resulting scalar potential in units of $M_{\rm Pl}/(4\pi)$ reads
\eq{
\label{scalarpotentialexb}
V =\bigg(
\frac{(h s+ \tilde{\mathfrak f})^2}{ 16s\op \tau^3} 
- \frac{6 h q s -2 q \tilde{\mathfrak f} }{ 16 s\op \tau^2} 
- \frac{5 q^2}{ 48  s\op  \tau}  \bigg)
+ \frac{\theta^2}{16 s\op \tau^3} \op,
}
which only depends on $s$, $\tau$, and the linear combination of axions $\theta=h\op c+
q\op\rho$. One linear combination of axions is not stabilized by
\eqref{scalarpotentialexb}, but can receive a mass from non-perturbative
effects. Such ultra-light axions can become part of dark radiation
\cite{Cicoli:2012aq}.

In \cite{Blumenhagen:2014nba,Blumenhagen:new} a mechanism to realize axion inflation 
together with moduli stabilization in string theory has been proposed. 
There the idea was to first stabilize all moduli except one axion
by turning on fluxes proportional to a large
parameter $\lambda$, and in a
second step stabilize the massless axion by
introducing a deformation depending on additional 
small fluxes.
The resulting superpotential takes the schematic form
\eq{
W_{\rm ax}=\lambda\op W+{f_{\rm ax}}\, \Delta W\,.
}
Instead of analyzing the resulting potential for the rather
complicated fully fledged models presented in \cite{Blumenhagen:new},
in this letter we mimic the resulting structure of the scalar potential by
introducing a flux parameter $\lambda$ in \eqref{scalarpotentialexb} as
\eq{
\label{scalarpotentialexa}
V =\lambda^2\bigg(
\frac{(h s+ \tilde{\mathfrak f})^2}{ 16s\op \tau^3} 
- \frac{6 h q s -2 q \tilde{\mathfrak f} }{ 16 s\op \tau^2} 
- \frac{5 q^2}{ 48  s\op  \tau}  \bigg)
+ \frac{\theta^2}{16 s\op \tau^3} \op.
}
We consider this as a (partly exactly solvable) toy model to analyze
the possibility of realizing large-field inflation in string theory.

Solving now \mbox{$\partial_s V=\partial_\tau V
=\partial_\theta V=0$,} we find three solutions.
Besides the  supersymmetric AdS minimum with a  tachyonic mode, there
exists a non-supersymmetric, tachyon-free AdS minimum at
\eq{
\label{minval}
  \tau_0={6 \op \tilde{\mathfrak f}\over 5\op q}\,,\hspace{20pt}
   s_0={\tilde{\mathfrak f}\over h}\,,\hspace{20pt}
   \theta_0=0\,.
}
To ensure $\tau_0, s_0 > 0$, for definiteness we chose all flux-quanta to be positive.
Furthermore, $\tilde{\mathfrak f}/h\gg 1$ and $\tilde{\mathfrak f}/q\gg 1$ implies
weak string coupling and large radius, so we can ignore higher-order
corrections to the scalar potential. The above fluxes also induce
D3- and a D7-brane tadpole charges, $N_{\rm D3}=-\lambda^2\tilde{\mathfrak f}\op h$ and $N_{\rm
D7}=-\lambda^2\tilde{\mathfrak f}\op q$, that need to be cancelled by D-branes.

To compute the mass eigenvalues and eigenstates in the
canonically-normalized basis, we consider
the matrix $M^i{}_j=K^{ik}\, V_{kj}$, with $V_{ij}={1\over
  2} \partial_i\partial_j V$, evaluated at the minimum.
We then find mass eigenvalues 
\eq{
\label{massemods}
M_{{\rm mod},i}^2=\mu_i \op{\lambda^2\, h \op q^3\over 16\op \tilde{\mathfrak f}^2} \, {M_{\rm Pl}^2\over 4\pi} \,,
}
with the numerical factors
\eq{
\mu_i=\left({\textstyle 0,{185\over 54\op\lambda^2}, {25\op(17-\sqrt{97})\over
         108},{25\op(17+\sqrt{97})\over 108}}\right).
}
The first two eigenstates are axionic while the last two are saxionic. In particular,
the massless state corresponds to an axionic linear combination.
Note that for sufficiently large $\lambda$, the axion can be parametrically lighter than the two saxions.

Since the volume and the dilaton are fixed by fluxes,
we  can  explicitly evaluate the string scale as
\eq{
\label{stringandKKscale}
   M_{\rm s}= {\sqrt{\pi}\op M_{\rm Pl}\over s^{1\over 4}\,(2\op\tau)^{3\over 4}}\,.
}
Recalling then \eqref{massemods} and \eqref{minval}, we derive the ratio
\eq{
\label{ratiomsmod}
     {M_{\rm s}\over M_{{\rm mod},i}}={13.03\over \sqrt{\mu_i}}{1\over
       \lambda\, h^{1\over 4} q^{3\over 4}}\,.
}


\section{Axion monodromy inflation}

\begin{figure}[t]
  \centering
  \vspace{0.4cm}
  \includegraphics[width=0.35\textwidth]{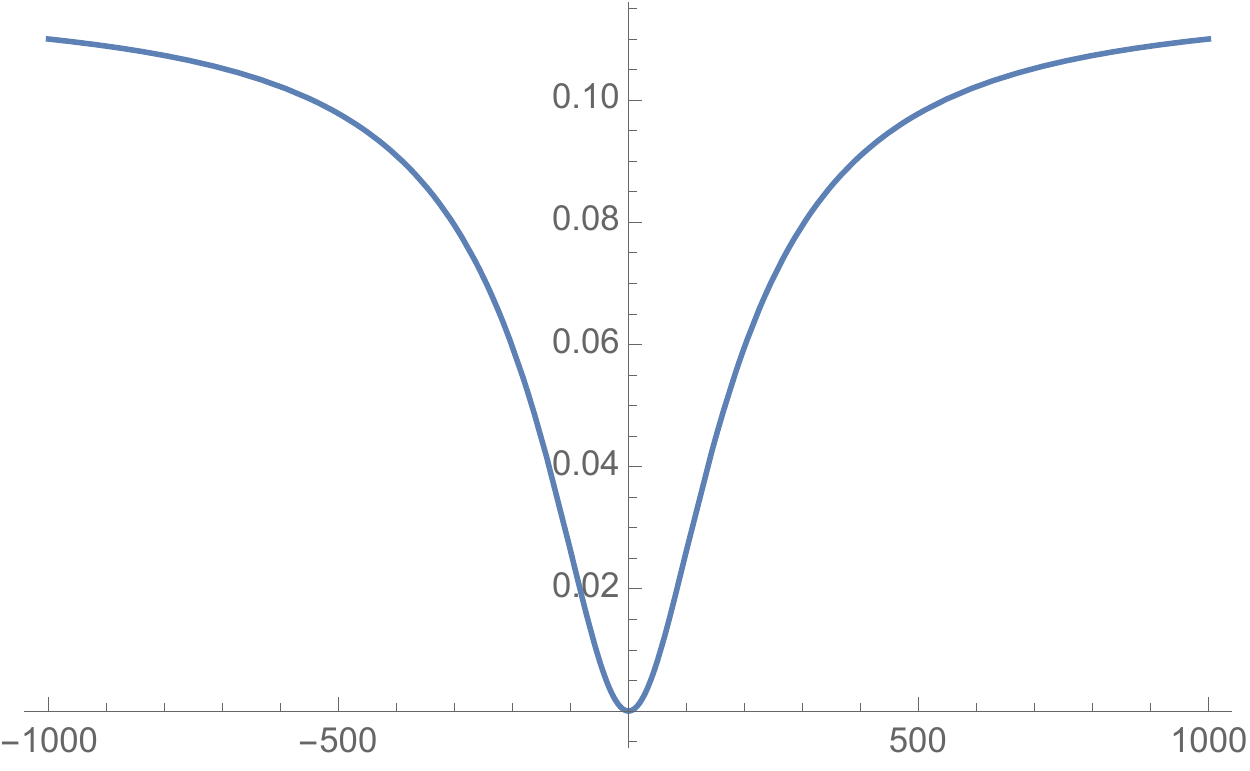}
  \begin{picture}(0,0)
  \put(0,6){\footnotesize$\theta$}
  \put(-102,116){\footnotesize $V_{\rm back}$}
  \end{picture}
  \caption{\small The potential $V_{\rm back}(\theta)$ shown in \eqref{backpotential} in units of
    $M^4_{\rm Pl}/(4\pi)$ for fluxes $h=1$, $q=1$, $\tilde{\mathfrak
    f}=10$ and $\lambda=10$. 
    For this large value of $\lambda$, the trajectory \eqref{backshift} correctly describes the 
    full motion in field space.
      }
  \label{bild1}
\end{figure}
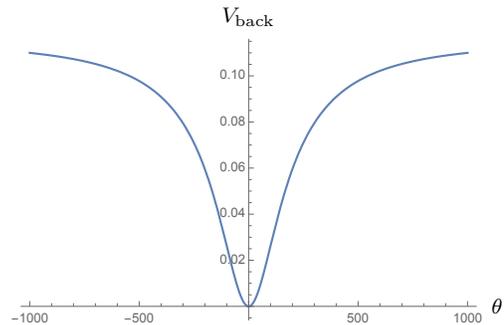

We now consider the backreaction effect of a 
slowly rolling and sufficiently light axion $\theta$, i.e. we take into account
that during the rolling the moduli $\tau$ and $s$ adjust 
adiabatically.  Solving 
the extremum conditions for a non-vanishing value of  $\theta$, we find
\eq{
\label{backshift}
    \tau_0(\theta)&={3\over 20\op q} \left(  4 \op\tilde{\mathfrak f}
      + \sqrt{10\op \left( \tfrac{\theta}{\lambda}\right)^2 +16\op
        \tilde{\mathfrak f}^2} \:\right),\\
    s_0(\theta)&={1\over 4\op h}  \,\sqrt{10\op \left( \tfrac{\theta}{\lambda}\right)^2 +16\op  \tilde{\mathfrak f}^2} \, .
}
For large $\lambda$, the motion in the full four-dimensional field space is well-described by \eqref{backshift};
for $\lambda$ of order one the trajectories differ, but qualitatively our results are still valid. 
Note also that for large excursions of $\theta$, the values of $\tau_0$ and $s_0$ 
are in
the perturbative regime, so that higher-order $\alpha'$- and
$g_s$-corrections to the scalar potential are under control.
Using \eqref{backshift} in
the potential \eqref{scalarpotentialexa} and performing a constant uplift  to vanishing cosmological
constant in the minimum, gives the following backreacted effective inflaton
potential (in units of $M_{\rm Pl}^2/4\pi$) 
\eq{
\label{backpotential}
  V_{\rm back}(\theta)=
  \frac{25 \lambda^2 h q^3}{108\op \tilde{\mathfrak f}^2} \op
  \frac{ 5 \bigl( \frac{\theta}{\lambda}\bigr)^2 - 4 \op\tilde{\mathfrak f}\op \Big(  4\op \tilde{\mathfrak f} - 
  \sqrt{
  10\bigl( \frac{\theta}{\lambda}\bigr)^2 +16\op \tilde{\mathfrak f}^2}  \Bigr)
  }{ 
  \Big(  4\op \tilde{\mathfrak f} + \sqrt{
    10\op\left( {\theta\over \lambda}\right)^2 +16\op \tilde{\mathfrak f}^2
  }  \Big)^2} .
}
Note that the initial simple quadratic potential is changed;
the expected flattening of the potential
becomes  evident in figure \ref{bild1}.
For small values of $\theta$ the potential still takes a quadratic form,
whereas for large  values of $\theta$ it becomes hyperbolic.
In the intermediate regime there is a turning point, around which  the
potential is linear.  
We remark that a non-constant uplift term of the form 
$V_{\rm up}=\varepsilon/\tau^\beta$ with $\beta$ small is also possible. 

Computing the mass  eigenvalues for each value of $\theta$, we find
that for large $\theta$ the eigenvalue along the trajectory becomes
tachyonic while the two transversal ones are positive.
Note also that for $\lambda \gg 1$, the mass hierarchy \eqref{massemods} remains
intact for $\theta\ne 0$.

Let us now analyze the potential  for the canonically-normalized 
inflaton in more detail.
For  $\theta/\lambda\ll \tilde{\mathfrak f}$ the shift in the value of the
minimum \eqref{backshift} is small and the potential takes 
a quadratic form 
\eq{
  \label{back_01}
     V_{\rm back}(\theta)\approx {125\op h\op q^3\over 3456 \op\tilde{\mathfrak f}^4}\, \theta^2\,.
}
Employing \eqref{backshift}, the total kinetic energy  
\eq{
\label{kinetics}
      {\cal L}_{\rm kin}=
      3\left({\partial \tau\over 2\op\tau}\right)^2
      +\left({\partial s\over 2\op s}\right)^2
      +3\left({\partial \rho\over 2\op\tau}\right)^2
      +\left({\partial c\over 2\op s}\right)^2 ,
}
determines the kinetic term for $\theta$.
To find the latter, we need to determine for every value
of $\theta$ the orthogonal combination $\sigma$ of axions. 
This can be fixed  as  
\eq{
\label{axbasis}
\partial \theta = h\op \partial c + q\op \partial\rho \,,\qquad
\partial \sigma = -\frac{q}{s^2}\op \partial c + \frac{3 h}{\tau^2} \op\partial \rho \,,
}
so that the axionic terms in \eqref{kinetics} become
\eq{ 
\label{axkin}
{\mathcal L}_{\rm kin}^{\rm ax} = 
\frac{3(\partial \theta)^2 + \tau^2 s^2 (\partial \sigma)^2}
{4(3h^2 s^2 + q^2 \tau^2)} \,.
}
For small $\theta$ this 
leads to a $\theta$-dependence of the form
${\cal L}_{\rm kin} \approx {25\over 148  \tilde{\mathfrak f}^2} (\partial \theta )^2$,
and thus the canonically-normalized inflaton takes the form $\Theta\approx
\sqrt{25/ 74}\,  {\theta/ \tilde{\mathfrak f}}$.
Note also that $\theta/\lambda\ll \tilde{\mathfrak f}$ implies $\Theta\ll \lambda$.

Next, we consider the large-field regime  $\theta/\lambda\gg \tilde{\mathfrak  f}$. 
Here we expand the backreacted potential \eqref{backpotential} as 
\eq{
    V_{\rm back}(\theta)\approx {25\over 216} \op {h\op q^3\op \lambda^2\over \tilde{\mathfrak f}^2} - {20\over 27}\op
    {h\op q^3\op \lambda^4\over \theta^2}\,.
}
We also approximate \eqref{backshift} by $\tau_0(\theta)\approx {3 \over 2 \sqrt{10}\op q\op \lambda} \theta$ and  
$s_0(\theta)\approx{\sqrt{10}\over 4\op h\op\lambda}  \theta$. 
Then, taking into account all fields from \eqref{kinetics}, we derive
\eq{
\label{fullkin}
  {\cal L}_{\rm kin}\approx 
  {2\over \gamma^2}\left( {\partial\theta\over \theta}\right)^2
  + \frac{15  }{896 h^2 q^2 \lambda^2} \theta^2 (\partial  \sigma)^2 ,
}
with $\gamma^2=28/(14 + 5\lambda^2)$.
Note that $\gamma$ is independent of the fluxes, but it depends on $\lambda$.
It can also be shown that, for appropriate initial conditions,
$\partial \sigma$ vanishes along the trajectory.
Canonically normalizing the inflaton via
\eq{
\theta=4 \op\sqrt{2\over 5}\, \tilde{\mathfrak f}\, \lambda\,\exp\left({\gamma\over 2}\op\Theta\right),
}
the potential in the large-field regime becomes Starobinsky-like
\eq{
  \label{back_02}
      V_{\rm back}(\Theta)={25\over 216}\op {h \op q^3\op \lambda^2\over \tilde{\mathfrak f}^2}
      \Big(1-e^{-\gamma\op\Theta}\Big)\,.
}
We emphasize that in contrast to the potential \eqref{back_01} in the small-field regime, 
the potential \eqref{back_02} is exponential due to the backreaction.

In the intermediate regime $\theta/\lambda \approx 1$, it is not possible to  take the canonical
normalization into account analytically. However, we show below that the full backreacted 
potential interpolates between a quadratic and a Starobinsky-like form.


\section{Qualitative Picture of Inflation}

Let us discuss how inflation can take place in this set-up, and how
its features depend on the value of $\lambda$. Here our
intuition has to be  based on the potential $V(\theta)$ for the non canonically
normalized potential, whereas a more accurate computation is presented in section \ref{sec_num}.


\paragraph{Quadratic inflation:}

\begin{figure}[t]
  \centering
  \vspace{0.4cm}
  \includegraphics[width=0.275\textwidth]{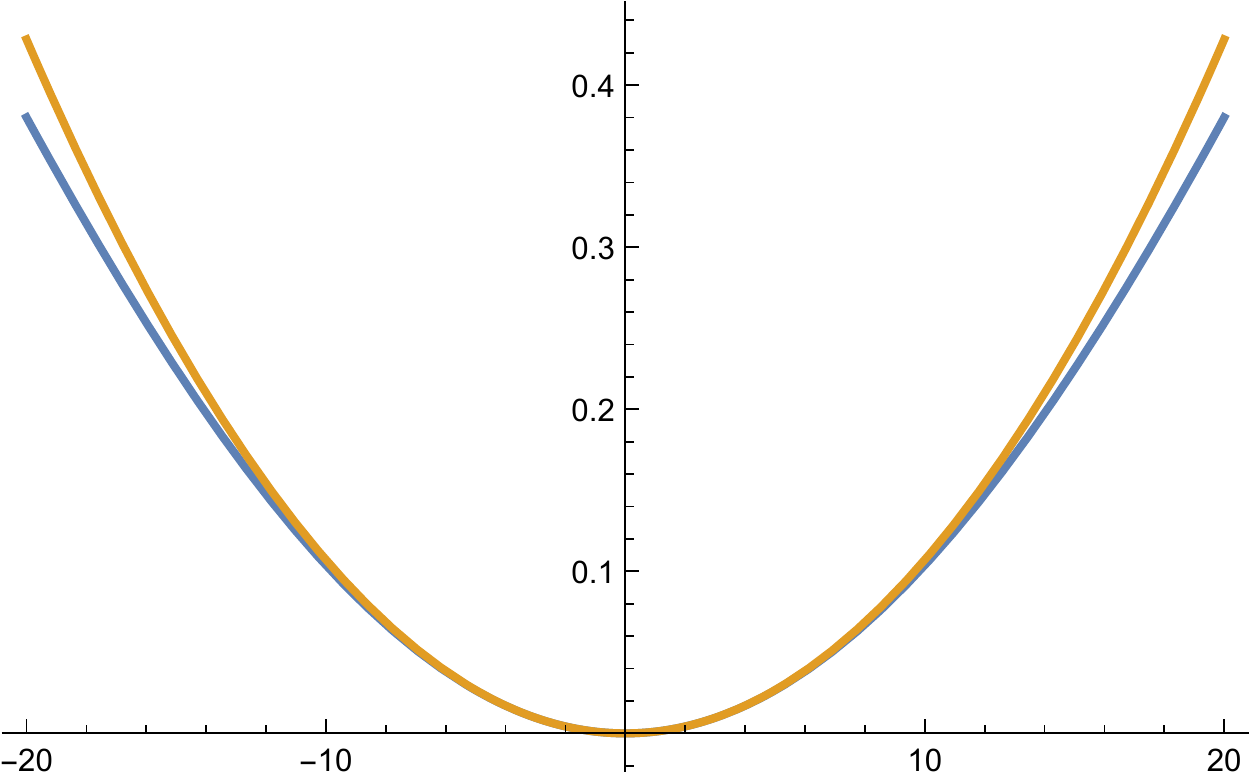}
  \begin{picture}(0,0)
    \put(0,4){\scriptsize$\Theta$}
    \put(-83,92){\scriptsize $V(\Theta)$}
  \end{picture}
  \caption{\small The potentials $V_{\rm back}(\Theta)$ and \eqref{back_01} in units of
    $M^4_{\rm Pl}/(4\pi)$ for fluxes $h=1$, $q=1$, $\tilde{\mathfrak
      f}=10$ and $\lambda=60$. The lower (blue) curve is the exact
    backreacted potential.}
  \label{bild4}
\end{figure}
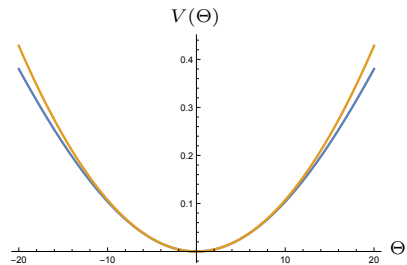

For sufficiently large $\lambda$ the backreacted potential \eqref{backpotential} is well
approximated by the quadratic term for the region $0<\Theta<15$, i.e.
slowly rolling down the potential one collects 60 e-foldings.
As is illustrated in figure \ref{bild4}, this is expected to happen for $\lambda \gtrsim 60$.
In this case, the inflaton is the lightest state and the heavy moduli
having masses of the order $M_{\rm mod}\sim \lambda\op M_{\Theta}$, which is larger
than the Hubble scale \mbox{$H\sim \sqrt{2 N_e/ 3}\op M_{\Theta}\sim 6.32\, M_{\Theta}$}.
Therefore, we have a model of single-field inflation and  all 
predictions agree with the ones of chaotic inflation,
in particular $r\sim 0.133$.
However, for such a large value of $\lambda$, the relation \eqref{ratiomsmod}
implies that the string scale becomes smaller than the heavy moduli
masses. Therefore, from the UV-complete point of view, using the
effective supergravity approximation
becomes questionable.

\begin{figure}[h]
  \centering
  \vspace{0.4cm}
  \includegraphics[width=0.275\textwidth]{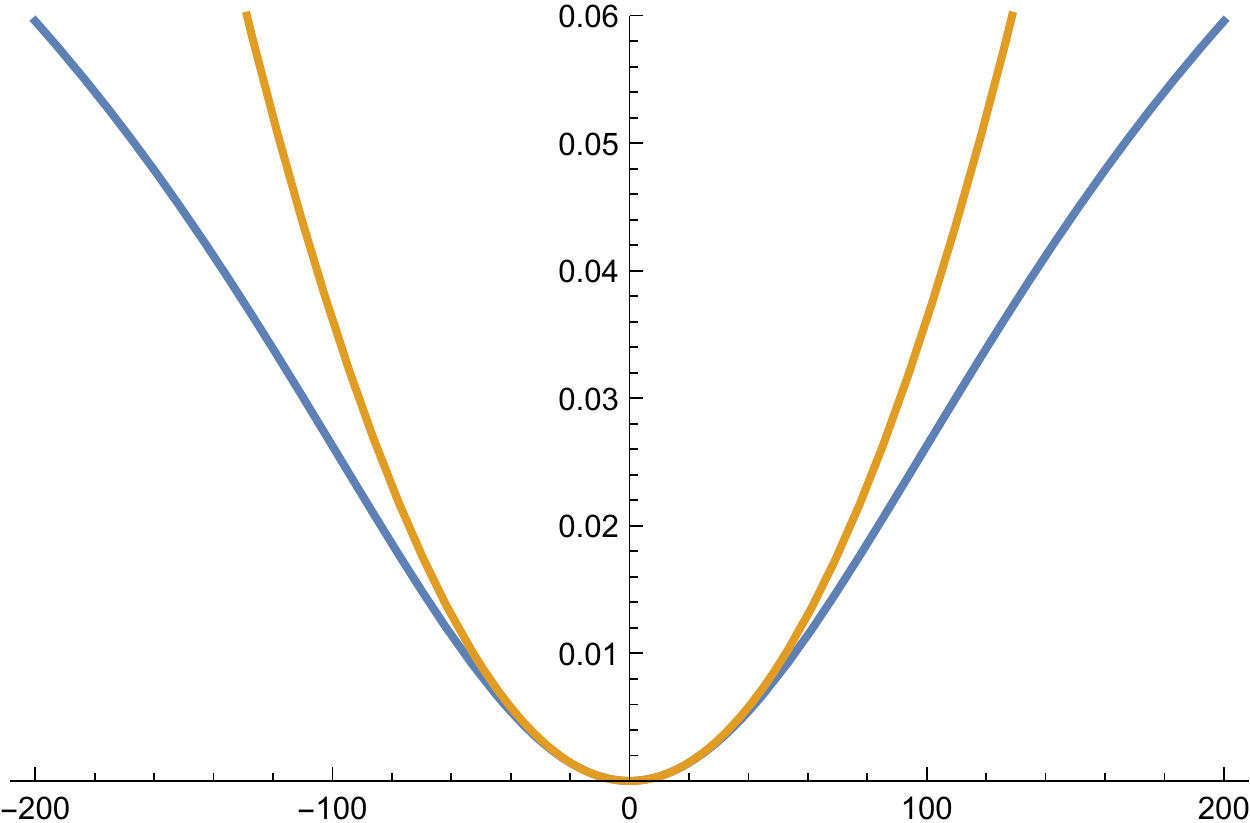}
  \begin{picture}(0,0)
    \put(0,3.5){\scriptsize$\theta$}
    \put(-81,97){\scriptsize $V(\theta)$}
  \end{picture}
  \caption{\small The potentials $V_{\rm back}(\theta)$ and  \eqref{back_01} in units of
    $M^4_{\rm Pl}/(4\pi)$ for fluxes $h=1$, $q=1$, $\tilde{\mathfrak
      f}=10$ and $\lambda=10$. }
  \label{bild3}
\end{figure}
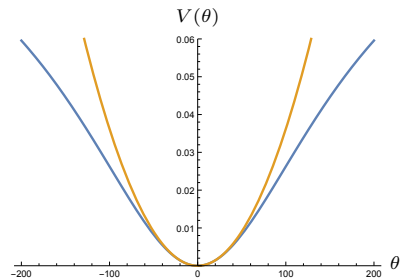

\paragraph{Linear inflation:} Lowering the value of $\lambda$, the
non-trivial  backreaction becomes more and more relevant,  that is the potential
becomes flatter in the large-field region. For $\lambda=10$ the full
potential
and the quadratic approximation are shown in figure \ref{bild3}.
Thus it is expected that most of the 50-60 e-foldings occur along  the
approximately linear potential. A more precise computation would require
the determination of the canonically normalized inflaton.
However, the expectation is that the tensor-to-scalar ratio becomes
smaller, namely $r\sim 0.08$ for linear inflation. The tension between
the string  scale and the heavy moduli masses becomes weaker,
while the heavy masses come closer to the Hubble scale.


\paragraph{Starobinsky-like  inflation:} 

\begin{figure}[h]
  \centering
  \vspace{0.4cm}
  \includegraphics[width=0.275\textwidth]{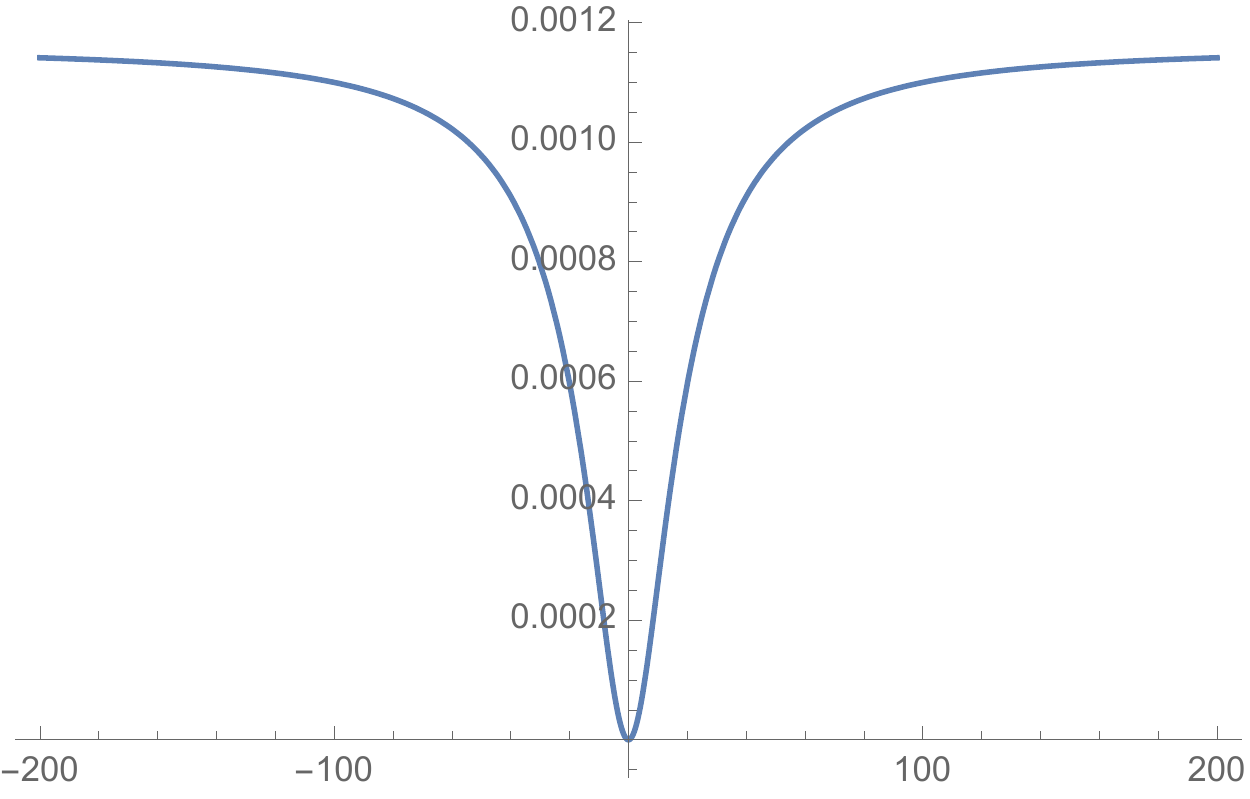}
  \begin{picture}(0,0)
    \put(0,5){\scriptsize$\theta$}
    \put(-89,96){\scriptsize $V_{\rm back}(\theta)$}
  \end{picture}
  \caption{\small The potential $V_{\rm back}(\theta)$  in units of
    $M^4_{\rm Pl}/(4\pi)$ for fluxes $h=1$, $q=1$, $\tilde{\mathfrak
      f}=10$ and $\lambda=1$. }
  \label{bild5}
\end{figure}
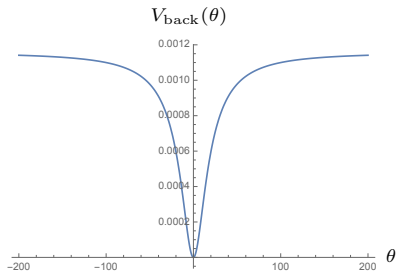

As figure \ref{bild5} shows, for $\lambda=O(1)$ the number of e-foldings
mainly would occur on the
Starobinsky-like plateau. 
In this case the tensor-to-scalar ratio becomes even
smaller and approaches the value  \mbox{$r\sim 0.0015$}. 
However, even though the heavy-moduli masses are lighter than the
string scale, they are now even  lighter than the Hubble scale.
Therefore, this is not  a model of single-field inflation
and the discussion of the inflationary trajectory becomes more
involved.


\section{Numerical analysis} 
\label{sec_num}

In this section, we numerically evaluate the tensor-to-scalar ratio
and the number of e-foldings, taking also the kinetic term into
account. We  see that qualitatively, the intuition from the
previous section is confirmed.

Starting from the kinetic terms \eqref{kinetics}, we can determine an effective  Lagrangian for
the field $\theta$ of the form
\eq{
   {\cal L}= \frac12 f(\theta)\,(\partial \theta)^2 +
   V(\theta)\, .
}
Expressing the Lagrangian in terms of the
canonically-normalized field is not always possible analytically, but one can
determine the slow-roll parameters also in terms of $\theta$ via
\eq{
\epsilon={1\over 2\op f}\left({V^\prime\over V}\right)^2\,,\qquad
  \eta={V^{\prime\prime}\over f\op V}- {f^\prime\op V^\prime\over 2 \op f^2\op V}
 \,,
}
where the prime denotes the  derivative with respect to $\theta$.
The number of e-foldings is given by
\eq{  N_e=\int_{\theta_{\rm end}}^{\theta^*} d\theta\, {f V\over 
    V^\prime} \, .
}

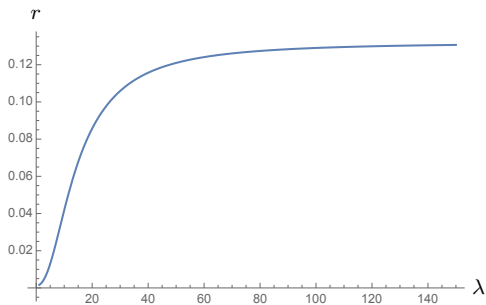
\begin{figure}[h]
  \centering
  \vspace{0.4cm}
  \includegraphics[width=0.34\textwidth]{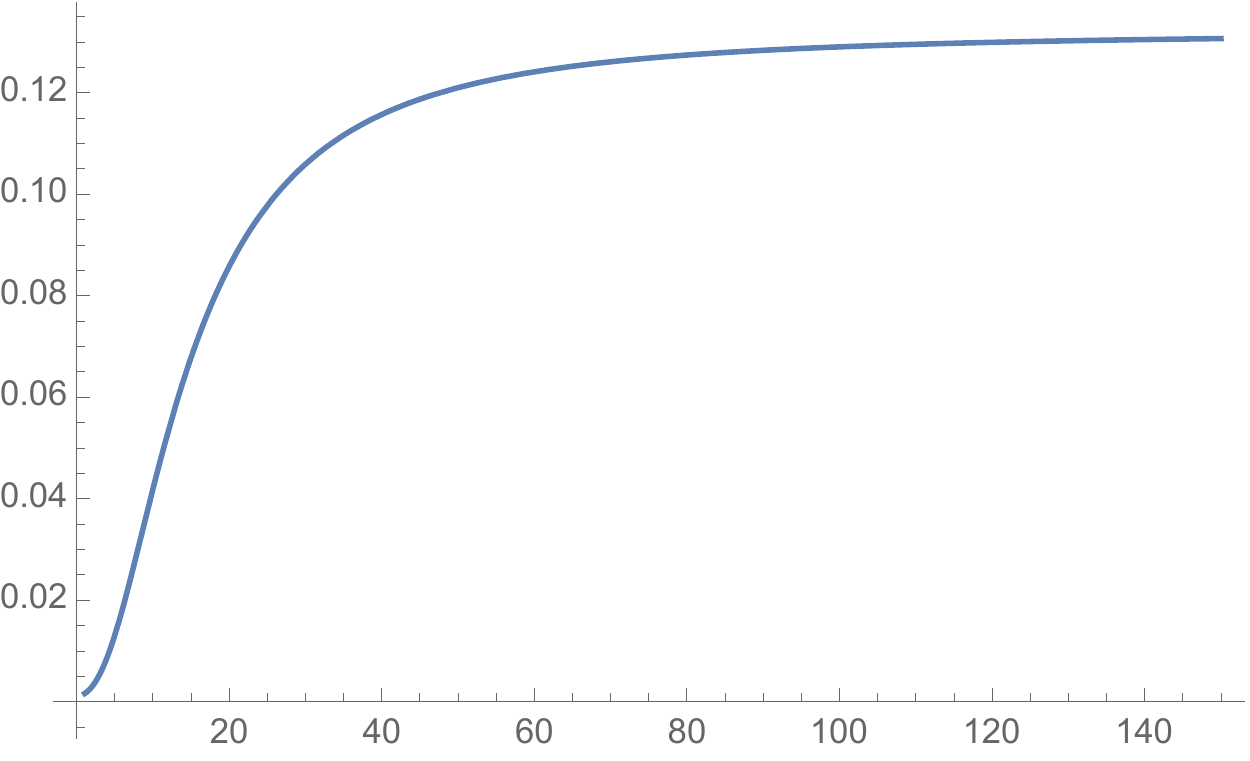}
  \begin{picture}(0,0)
    \put(-1,6){\footnotesize$\lambda$}
    \put(-168,110){\footnotesize $r$}
  \end{picture}
  \caption{\small The tensor-to-scalar ratio as a function of
    $\lambda$ for fixed $n_s=0.967$.}
  \label{bild6}
\end{figure}

\begin{figure}[h]
  \centering
  \vspace{0.5cm}
  \includegraphics[width=0.34\textwidth]{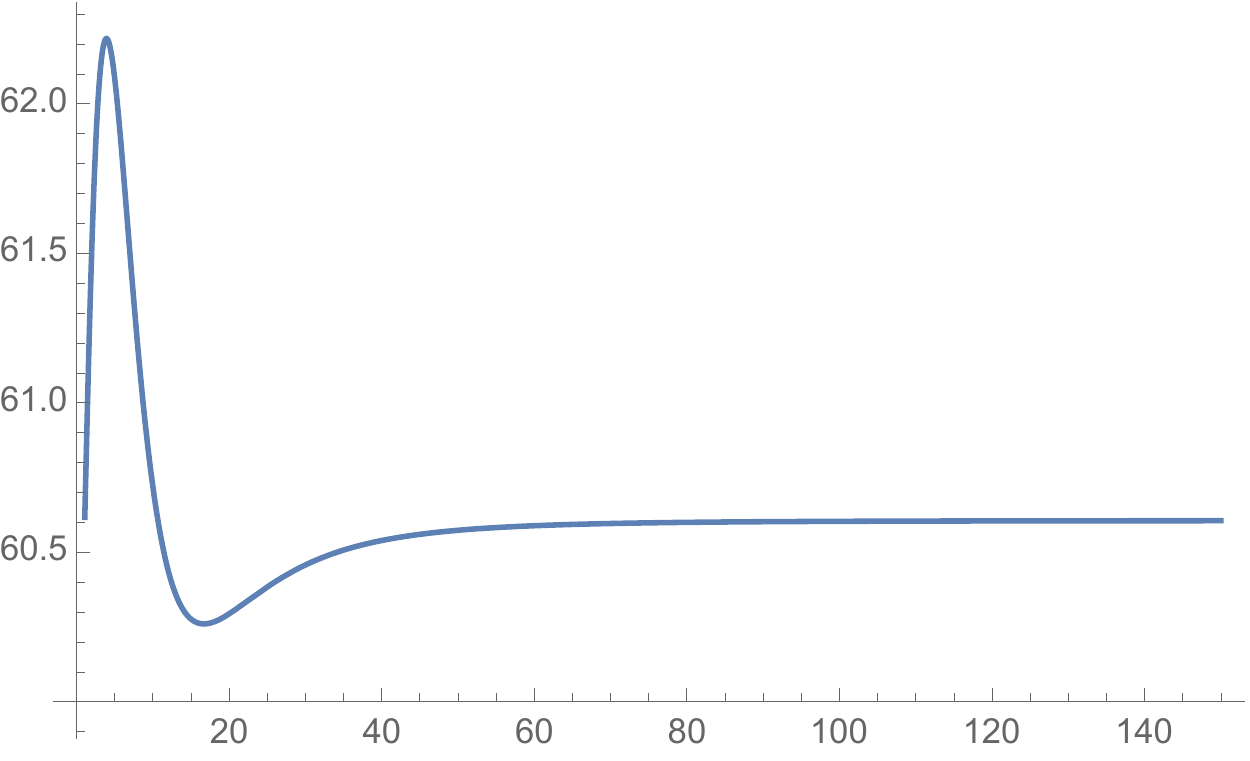}
  \begin{picture}(0,0)
    \put(-1,6){\footnotesize$\lambda$}
    \put(-171,111){\footnotesize $N_{\rm e}$}
  \end{picture}
  \caption{\small 
The number of e-foldings  as a function of
    $\lambda$ for fixed $n_s=0.967$.}
  \label{bild7}
\end{figure}
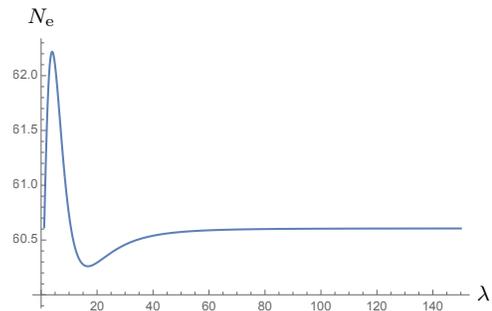

To evaluate $f(\theta)$ we substitute \eqref{backshift} and \eqref{axkin}
in \eqref{kinetics}.
We can then numerically determine the tensor-to-scalar ratio
in terms of $\lambda$ (for fixed fluxes) by fixing $n_s=0.967$.
The resulting behavior is displayed in figure \ref{bild6}, whereas 
figure \ref{bild7} shows the corresponding number of e-foldings.

The curves show the expected behavior, 
namely that with decreasing $\lambda$ the model changes from chaotic to Starobinsky-like
inflation.


\section{Conclusions}

In a simple string compactification with two complex moduli, after
introducing by hand a scaling parameter $\lambda$, 
we were able to stabilize all moduli except a single axion using 
NS-NS and R-R three-form flux together with non-geometric $Q$-flux. 
The hierarchically-light but massive axion served as an inflaton candidate.
Taking into account the backreaction and assuming an uplift to Minkowski, 
we evaluated the resulting potential, which turned
out to interpolate between a quadratic and a Starobinsky-like potential.
We analyzed the cosmological consequences for
three different regimes of $\lambda$. 
Depending on $\lambda$,
the tensor-to-scalar ratio  interpolates between the one
for chaotic and the one for Starobinsky-like inflation.

From a controllable UV-complete theory point of view, large-field inflation models 
require a hierarchy of the form
\eq{
\label{scales}
M_{\rm Pl}> M_{\rm s}>M_{\rm KK}> M_{\rm mod} >H_{\rm inf} > M_{\Theta}\,,
}
where neighboring scales can differ by (only) a factor of $O(10)$. 
Our main observation was: the larger $\lambda$, the more difficult it becomes to separate the high scales on
the left of \eqref{scales}.  
Contrarily, for small $\lambda$, the smaller (Hubble-related) scales on
the right of  \eqref{scales} become difficult to separate.

Corrections to the scalar potential are expected to be under control,
due to the shift symmetry of the axion/inflaton and due to the
adiabatic adjustment of the saxionic moduli into the perturbative
regime. 
Our string-motivated analysis  shows how a Starobinsky-like inflation
model could arise  from string-theory axions. For a more realistic scenario, however,
more string model building is needed,
including the introduction of an MSSM-like D7-brane set-up and  the 
computation of soft-supersymmetry breaking terms \cite{Blumenhagen:new}.
Note also that since the inflaton in our model is a linear combination of the universal axion and
a K\"ahler axion, we can realize the stringy reheating mechanism proposed in
\cite{Blumenhagen:2014gta}.


\vspace{1.6pt}

\noindent
\emph{Acknowledgments:}
We thank G. Dall'Agata, F. Quevedo, G. Raffelt and T. Weigand for
discussions. We are particularly grateful to F. Pedro for pointing
out an important subtlety in an earlier version of this letter.
R.B. thanks the Charles University in Prague for hospitality. 
A.F. thanks the AvH Foundation for support.
E.P. is supported by the MIUR grant FIRB RBFR10QS5J
and by the Padua University Project CPDA144437.



\end{document}